%% file: biermann_beuther.tex
\documentclass[twoside]{article}
\usepackage[latin1]{inputenc}
\usepackage{t1enc}
\usepackage{a4}
\usepackage{tabularx}
\usepackage{epsf}
\usepackage{graphicx}
\usepackage{natbib}
\bibpunct{(}{)}{;}{a}{}{,}

 \renewenvironment{thebibliography}[1]{%
   \begin{oldthebibliography}{#1}%
     \setlength{\parskip}{0ex}%
     \setlength{\itemsep}{0ex}%
 }%
 {%
   \end{oldthebibliography}%
 }

\begin{document}

\setcounter{figure}{0}
\setcounter{section}{0}
\setcounter{equation}{0}

\begin{center}
{\Large\bf
Massive Star Formation:\\[0.2cm]
The Power of Interferometry}\\[0.7cm]

Henrik Beuther \\[0.17cm]
Max Planck Institute for Astronomy \\
K\"onigstuhl 17, 69117 Heidelberg, Germany \\
beuther@mpia.de
\end{center}

\vspace{0.5cm}

\begin{abstract}
\noindent{\it
  This article presents recent work to constrain the physical and
  chemical properties in high-mass star formation based largely on
  interferometric high-spatial-resolution continuum and spectral line
  studies at (sub)mm wavelengths. After outlining the concepts,
  potential observational tests, a proposed evolutionary sequence and
  different possible definitions for massive protostars, four
  particular topics are highlighted: (a) What are the physical
  conditions at the onset of massive star formation? (b) What are the
  characteristics of potential massive accretion disks and what do
  they tell us about massive star formation in general? (c) How do
  massive clumps fragment, and what does it imply to high-mass star
  formation? (d) What do we learn from imaging spectral line surveys
  with respect to the chemistry itself as well as for utilizing
  molecules as tools for astrophysical investigations?}
\end{abstract}

\section{Introduction}
\label{intro}

Star formation is a key process in the universe, shaping the structure
of entire galaxies and determining the route to planet formation. In
particular, the formation of massive stars impacts the dynamical,
thermal and chemical structure of the interstellar medium (ISM), and
it is almost the only mode of star formation observable in
extragalactic systems. During their early formation, massive
star-forming regions inject energy to the ISM via their outflows and
jets, then during their main sequence evolution, the intense
uv-radiation of the massive stars heats up their environment, and at
the very end of their life, Supernovae stir up the ISM by their
strong explosions. Furthermore, massive stars are the cradles of the
heavy elements. Hence, life as we know it today would not exist if
massive stars had not formed first the heavy elements in their
interiors via nucleosynthesis. In addition to this, almost all
stars form in clusters, and within the clusters massive stars dominate
the overall luminosities. Isolated low-mass star formation is the
exception, and isolated high-mass star formation likely does not exist

{\it Concepts:} In spite of their importance, many physical processes
during the formation of massive stars are not well understood. While
there exists a paradigm for low-mass star formation on which large
parts of the scientific community agree (e.g.,
\citealt{andre2000,mckee2007}), this is less the case for high-mass
star formation (e.g., \citealt{beuther2006b,zinnecker2007}). The
conceptional problem is based on the fact that at least in
spherical symmetry the radiation pressure of a centrally ignited star
$\geq$8\,M$_{\odot}$ would be large enough to stop any further gas
infall and hence inhibit the formation of more massive objects. Over
the last decade, two schools have been followed to solve this problem:
(a) The turbulent accretion scenario, which is largely an enhancement
of the low-mass star formation scenario, forms massive stars in a
turbulent core with high accretion rates and a geometry including
accretion disks and molecular outflows (e.g.,
\citealt{yorke2002,mckee2003,krumholz2006b,keto2007}). In contrast to
that, (b) the competitive accretion scenario relies on the clustered
mode of massive star formation.  In this scenario, the accretion rates
are determined by the whole cluster potential, and those sources
sitting closest to the potential well will competitively accrete most
of the mass (e.g., \citealt{bonnell2004,bonnell2006}).

{\it Potential observational tests:} How can we observationally
discriminate between the different scenarios? For example, the
turbulent accretion scenario predicts qualitatively similar outflow
and disk properties as known for low-mass stars, however, with
quantitatively enhanced parameters like the accretion rates, outflow
energies, or disk sizes.  In contrast to that, modeling of the
proto-cluster evolution in the competitive accretion scenario
indicates extremely dynamic movements of all cluster-members
throughout their whole evolution. It is unlikely that in such a
dynamic environment collimated outflows or large disks could survive
at all. Another difference between both scenarios is based on their
early fragmentation predictions. While the turbulent accretion
scenario draws their initial gas clumps from (gravo-)turbulently
fragmented clouds (e.g., \citealt{padoan2002}) and these gas clumps do
not fragment afterwards much anymore, the competitive accretion
scenario predicts that the initial gas clumps fragment down to many
clumps all of the order a Jeans-mass ($\sim$0.5\,M$_{\odot}$). Hence,
while in the former scenario, the Initial Mass Function (IMF) is
determined during the early cloud fragmentation processes, the latter
models predict that the IMF only develops during the ongoing cluster
formation. Therefore, studying the initial fragmentation and the early
core mass functions can give insights in the actual massive
star formation processes.

{\it Evolutionary sequence:} Independent of the formation scenario,
there has to be an evolutionary sequence in which the processes take
place. For this review, I will follow the evolutionary sequence
outlined by \citet{beuther2006b}: Massive star-forming regions start
as High-Mass Starless Cores (HMSCs), i.e., these are massive gas cores
of the order a few 100 to a few 1000\,M$_{\odot}$ without any embedded
protostars yet. In the next stage we have High-Mass cores with
embedded low- to intermediate-mass protostars below 8\,M$_{\odot}$,
which have not started hydrogen burning yet.  During that evolutionary
phase, their luminosity should still be dominated by accretion
luminosity. Following that, the so called High-Mass Protostellar
Objects (HMPOs) are still massive gas cores but now they contain
embedded massive protostars $>$8\,M$_{\odot}$ that have started
hydrogen burning which soon dominates the total luminosity of the
sources. Hot Molecular Cores (HMCs) and hypercompact H{\sc ii} regions
(HCH{\sc ii}s) are part of that class. The last evolutionary stage
then contains the final stars that have stopped accreting. While most
ultracompact H{\sc ii} regions (UCH{\sc ii}s) are likely part of the
latter group, some of them may still be in the accretion phase and
could then hence still harbor HMPOs.

{\it Definitions of a massive protostar:} Another debate in massive
star formation centers around the exact definition of a "massive
protostar". If one followed the low-mass definition which basically
means that a protostars is an object that derives most of its
luminosity from accretion, then "massive protostars" should not exist
or only during a very short period of time because as soon as they are
termed "massive" ($>$8\,M$_{\odot}$), their luminosity is quickly
dominated by hydrogen burning. In this scenario, during the ongoing
formation processes, one would then need to talk about "accreting
stars". This approach is for example outlined by
\citet{zinnecker2007}. A different definition for "massive protostars"
is advocated, e.g., recently by \citet{beuther2006b}: In this picture,
a protostar is defined in the sense that each massive object that is
still in its accretion phase is called a "massive protostar",
independent of the dominating source of luminosity. This definition
follows more closely the usual terminology of "proto", meaning objects
that are not finished yet.

{\it Observational challenges:} Whatever physical or chemical
processes we are interested in massive star formation, one faces
severe observational challenges because of the clustered mode of
massive star formation and the on average large distances of a few
kiloparsec. Therefore, high spatial resolution is a prerequisite for
any such study. Furthermore, the early stages of massive star
formation are characterized by on average cold gas and dust
temperatures which are best observed at (sub)mm wavelength. Hence,
most observations presented in the following are based on (sub)mm
interferometer observations of young massive star-forming regions at
different evolutionary stages.

The main body of this article is divided into four sections dealing
first with the initial conditions that are present prior to or at the
onset of massive star formation. The next two sections deal with our
current knowledge about the properties of potential massive accretion
disks and the fragmentation behavior of massive gas clumps and cores.
The following section will then outline the status and future
possibilities of astrochemical investigations. Finally, I try to
sketch the directions where current and future answers to the
questions raised in the Abstract may lead to.

\section{The earliest stages of massive star formation}
\label{early}

What are the initial conditions of massive star formation? Until a few
years ago, to address this question observationally in a statistical
sense was close to impossible because we had no means to identify
large samples of sources prior to or at the onset of massive star
formation. The situation has changed significantly since the advent of
the space-based near- and mid-infrared missions that surveyed the
Galactic plane starting with ISO and MSX, and now conducted with much
better sensitivity and spatial resolution by Spitzer. These missions
have revealed more than $10^4$ Infrared Dark Clouds (IRDCs), which are
cold molecular clouds that are identified as shadows against the
Galactic background (e.g., \citealt{egan1998,carey2000,simon2006}).
These clouds are characterized by on average cold temperatures
($\sim$15\,K), large masses (a few 100 to a few 1000\,M$_{\odot}$) and
average densities of the order $10^4-10^5$\,cm$^{-1}$ (e.g.,
\citealt{rathborne2005,sridharan2005,pillai2006}). Although these
clouds appear as dark shadows, they may be starless but they can also
harbor embedded forming protostars. In fact, a statistical analysis
of the percentage of starless IRDCs versus IRDCs with embedded
protostars will be an important step to understand the time-scales
important for the earliest evolutionary stages. Currently, the
statistical database of in depth IRDC studies is still insufficient
for such an estimate (e.g.,
\citealt{rathborne2005,rathborne2006,pillai2006b,beuther2007a,motte2007}),
however, it is interesting to note that until now no starless IRDC has
been unambiguously identified in the literature, all detailed studies
revealed embedded star formation processes. To first order, this
triggers the interpretation/speculation that the high-mass starless
core phase is likely to be extremely shortlived. Future investigations
of larger samples will answer this question more thoroughly.

\begin{figure}[htb] \begin{center}
\end{center}
\caption{\small Sample IRDCs from \citet{sridharan2005}: MSX A-band
  (8\,$\mu$m) images (black is bright) with 1.2\,mm emission contours:
  The first two numbers refer to the corresponding IRAS source and the
  third number labels the mm sub-sources. The five-pointed stars mark
  cores lacking good 1.2 mm measurements.}  \label{irdcs}
\end{figure}

In a recent spectral line study of a sample of 43 IRDCs
(Fig.~\ref{irdcs}), \citet{beuther2007g} detected SiO(2--1) emission
from 18 sources.  Assuming that SiO is produced solely through
sputtering from dust grains, and that this sample is representative
for IRDCs in general, it indicates that at least 40\% of the IRDCs
have ongoing outflow activity. Since the non-detection of SiO does not
imply no outflow activity, this number is a lower limit, and even a
higher percentage of sources may harbor already ongoing star
formation. The range of observed SiO line-widths down to zero
intensity varied between 2.2 and 65\,km\,s$^{-1}$. While inclination
effects and embedded objects of different mass could account for some
of the differences, such effects are unlikely causing the whole
velocity spread. Therefore, \citet{beuther2007g} speculate whether the
varying SiO line-widths are also indicators of their evolutionary
stage with the smallest line-width close after the onset of star
formation activity. In the same study, \citet{beuther2007g} observed
CH$_3$OH and CH$_3$CN. While CH$_3$CN was detected only toward six
sources, CH$_3$OH was found in approximately 40\% of the sample. The
derived column densities are low of the order $10^{-10}$ with respect
to H$_2$.  These values are consistent with chemical models of the
earliest evolutionary stages of high-mass star formation (e.g.,
\citealt{nomura2004}), and the CH$_3$OH abundances compare well to
recently reported values for low-mass starless cores (e.g.,
\citealt{tafalla2006}).

Zooming into selected regions in more detail, we studied one
particularly interesting IRDC at high angular resolution with the
Plateau de Bure Interferometer and the Spitzer Space Telescope
(IRDC\,18223-3, see Fig.~\ref{irdcs} right panel;
\citealt{beuther2005d,beuther2007a}). Combining the Spitzer
mid-infrared data between 3 and 8\,$\mu$m with the 3.2\,mm
long-wavelengths observations from the Plateau de Bure Interferometer
(PdBI), we did not find any mid-infrared counterpart to the massive
gas core detected at 3.2\,mm (Fig.~\ref{18223-3},
\citealt{beuther2005d}). However, we did detect three faint
4.5\,$\mu$m features at the edge of the central 3.2\,mm continuum
core. Since emission features that occur only in the 4.5\,$\mu$m but
no other Spitzer band are usually attributed to shocked H$_2$ emission
from molecular outflows (e.g., \citealt{noriega2004}), we concluded
that the region likely hosts a very young protostar that drives a
molecular outflow but that is still too deeply embedded by to be
detected by the Spitzer IRAC bands. This interpretation found further
support by line-wing emission in older CO and CS data. Based on the
inferred central source, we predicted that the region should have a
strongly rising spectral energy distribution (SED) and hence be
detected at longer wavelengths.  As soon as the MIPSGAL mid- to
far-infrared survey with Spitzer became available, we then could
identify the central source at 24 and 70\,$\mu$m (Fig.~\ref{18223-3},
\citealt{beuther2007a}). Combing the available mid-/far-infrared data
with the long-wavelengths observations in the mm regime, it is
possible to fit the SED with a two component model: one cold component
($\sim$15\,K and $\sim$576\,M$_{\odot}$) that contains most of the
mass and luminosity, and one warmer component ($\sim$51\,K and
$\sim$0.01\,M$_{\odot}$) to explain the 24\,$\mu$m data. The
integrated luminosity of $\sim$177\,L$_{\odot}$ can be used to
constrain additional parameters of the embedded protostar from the
turbulent core accretion model for massive star formation
\citep{mckee2003}. Following the simulations by \citet{krumholz2006b},
the data of IRDC\,18223-3 are consistent with a massive gas core
harboring a low-mass protostellar seed of still less than half a solar
mass with high accretion rates of the order
$10^{-4}$\,M$_{\odot}$yr$^{-1}$ and an age below 1000\,yrs. In the
framework of this model, the embedded protostar is destined to become
a massive star at the end of its formation processes. While this
interpretation is attractive, it is not unambiguous, and especially
the derived time-scale from this model appears short when comparing
with recent outflow data that will be presented in the following
section (\S\ref{18223-cassie}).

\begin{figure}[htb]
\caption{\small IRDC\,18223-3 images at different wavelengths from
  \citet{beuther2007a}. The color scales show Spitzer images at
  various wavelength, and the contours show the 93\,GHz (3.2mm)
  continuum emission observed with the PdBI \citep{beuther2005d}. The
  left panel presents a three-color composite with blue 3.6\,$\mu$m,
  green 4.5\,$\mu$m and red 8.0\,$\mu$m (adapted from
  \citealt{beuther2005d}). The inlay zooms into the central core
  region. The middle and right panel show the Spitzer 24 and
  70\,$\mu$m images, respectively. The circles in each panel present
  the Spitzer beam sizes and the ellipse in the left panel presents
  the PdBI 3.2\,mm continuum synthesized beam. A size-ruler is also
  shown in the left panel.}  \label{18223-3} \end{figure}

In summary, these observations indicate that the physical and chemical
conditions at the onset of low- and high-mass star formation do not
differ significantly (except for largely different initial cloud clump
masses and accretion rates), and that the time-scale for massive bound
gas clumps to remain starless is likely relatively short.

\section{Massive accretion disks?}
\label{disks}

As mentioned in the Introduction, molecular outflows and accretion
disks can be used to discriminate between the different formation
scenarios for massive stars. Massive outflows have been subject to
intense studies for more than a decade (e.g.,
\citealt{shepherd1996a,beuther2002b,zhang2005,arce2006}), and although
there is considerable discussion about the details, we find a growing
consensus that massive molecular outflows are ubiquitous in high-mass
star formation, and that collimated jet-like outflows do exist for
massive sources as well, at least during the very early evolutionary
stages \citep{beuther2005b}. The collimation of the outflows is likely
to widen with ongoing evolution. Nevertheless, these data are
consistent with the turbulent core model for massive star formation,
whereas they are less easy to reconcile with the competitive accretion
model because the latter is so dynamic that collimated structures
likely could not survive very long. Furthermore, the existence of
collimated outflows can only be explained by magneto-hydrodynamic
acceleration of the jet from an underlying accretion disk. Hence,
there is ample indirect evidence for massive accretion disks, however,
the physical characterization of disks in massive star formation is
still lacking largely the observational basis \citep{cesaroni2006}.
The two main reasons for this are that the expected massive accretion
disks are still deeply embedded within their natal cores complicating
the differentiation of the disk emission from the ambient core, and
that the clustered mode of massive star formation combined with the
large average distances of the targets makes spatially resolving
structures of the order 1000\,AU a difficult observational task. In
spite of these difficulties, the advent of broad spectral bandpasses
allowing us to study several spectral lines simultaneously, as well as
the improved spatial resolution of existing and forthcoming
interferometers have increased the number of disk studies over the
last few years. For a recent review see \citet{cesaroni2006}.

Here, I will show three different examples of disk and/or rotation
candidates in an evolutionary sense: It starts with a rotation and
outflow investigation of the previously discussed IRDC\,18223-3
(\S\ref{early}), then I will present recent data from the high-mass disk
candidate in the HMPO IRAS\,18089-1732, and finally observations from
a massive disk candidate at a more evolved evolutionary stage will be
discussed.

\subsection{Rotation in IRDCs: the case of IRDC\,18223-3}
\label{18223-cassie}

As a follow-up of the Infrared Dark Cloud study of IRDC\,18233-3
discussed in \S\ref{early}, Fallscheer et al.~(in prep.) observed the
same region with the Submillimeter Array (SMA) in several spectral
setups around 230 and 280\,GHz covering outflow as well as dense gas
tracers. Figure \ref{cassie} shows a compilation of the CO(2--1) data,
one CH$_3$OH line and the dust continuum emission. The blue- and
red-shifted CO(2--1) emission clearly identifies at least one
large-scale outflow in the north-west south-east direction. This is
consistent with two of the 4.5\,$\mu$m emission features at the edge
of the core (Fig.~\ref{18223-3}, left panel and inlay). There is
another collimated red-shifted feature to the south-west corresponding
to the third 4.5\,$\mu$m feature, however, we do not identify a blue
counterpart and refrain from further interpretation of that feature.
Since we find for the main north-west south-east outflow blue- and
red-shifted emission on both sides of the continuum peak, the
orientation of the outflow should be close to the plane of the sky
(see, e.g., models by \citealt{cabrit1990}), and hence the assumed
underlying perpendicular rotating structure close to edge-on.
Following the approach outlined in \citet{beuther2002b}, the outflow
mass and outflow rate are 13\,M$_{\odot}$ and $3.5\times
10^{-4}$\,M$_{\odot}$yr$^{-1}$, respectively. With the above derived
core mass of $\sim$576\,M$_{\odot}$ (\S\ref{early}), this source fits
well into the correlation between outflow rate and core mass
previously derived for HMPOs (Fig.~7 in \citealt{beuther2002b}).

\begin{figure}[htb] 
\caption{\small SMA observations toward IRDC\,18223-3 (Fallscheer et
  al.~in prep.). The left panel shows the blue- and red-shifted
  CO(2--1) emission as solid and dashed contours overlaid on the
  grey-scale 1.3\,mm dust continuum emission. The central core is the
  same source as in Fig.~\ref{18223-3}. The right panel presents in
  grey-scale a 1st moment map (intensity weighted velocity
  distribution) of CH$_3$OH overlaid with the 1.1\,mm continuum
  emission. The empty and full circle are the synthesized beams of the
  line and continuum emission, respectively.} \label{cassie}
\end{figure}

While the outflow rate is consistent with the accretion rate
previously derived from the SED (\S\ref{early}), discrepancies arise
with respect to the age of the system. Although dynamical timescales
are highly uncertain (e.g., \citealt{parker1991}), the size of the
molecular outflow combined with a low inclination angle allows for at
least a timescale estimate for the outflow of the order a few
$10^4$\,yrs, well in excess of the value $\leq 10^3$\,yrs previously
derived from the SED applied to models (\S\ref{early}).
Notwithstanding the large errors between the different estimates, the
discrepancy of more than an order of magnitude appears real. How can
we explain that?  There is no clear answer to that yet, but a
possibility is that the orientation of the disk-outflow system with
the disk close to edge-on absorbs a large amount of flux distorting
the SED on the Wien-side. If that were the case, the SED-estimated age
could underestimate the age of the system. Another possibility
to solve the discrepancy is that the initial start of high-mass star
formation may proceed slower, i.e., the first low-mass protostar(s)
(destined to become massive or not?)  form within the massive cores,
and they already start driving outflows, but at that stage it is
impossible to detect them in the near- to far-infrared because of the
large extinction. In this picture at some point the high-mass star
formation process would need to accelerate because otherwise the
massive stars cannot form in the short time-scales of a few
$10^5$\,yrs (e.g., \citealt{mckee2002}). It is not clear why the whole
process should start slow and what could trigger such acceleration
later-on. Obviously, more theoretical and observational work is
required to explain the different time-scales in more detail.

Figure \ref{cassie} (right panel) zooms into the central core and
shows dust continuum emission as well as the velocity structure of the
dense central gas observed in CH$_3$OH$(6_{0,1}-5_{0,1})$ with a lower
level excitation level of $E_{\rm{low}}=34.8$\,K. Interestingly, both
the continuum and the spectral line emission are elongated in the
north-east south-west direction perpendicular to the main molecular
outflow.  While the continuum emission shows three resolved emission
features, CH$_3$OH exhibits a smooth velocity gradient across the
source spanning approximately 3\,km\,s$^{-1}$. The CH$_3$OH line-width
FWHM toward the continuum peak is 2.1\,km\,s$^{-1}$. The
blue-redshifted features in the north-west are likely part of the
molecular outflow and one sees even a slight elongation of the
continuum emission in that direction.  While CH$_3$OH is a well-known
shock tracer and hence regularly found within molecular outflows
(e.g., \citealt{bachiller2001}), it is more of a surprise that we find
it in an elongated structure likely associated with rotation and
infall perpendicular to the outflow. The extent of this structure is
large with $\sim$6.5$''$ corresponding to more than 20000\,AU at the
given distance of 3.7\,kpc. Although we have no methanol isotopologue
in the setup to exactly determine the opacity of the line, a
low-energy transition like this is likely to be optically thick.
Hence, we are tracing some of the outer rotating structures, probably
corresponding to a larger scale rotating and potentially
infalling/inspiralling toroid (e.g., \citealt{cesaroni2005,keto2007}).
The small velocity spread across the structure as well as the
relatively narrow CH$_3$OH line-width toward the core center are also
consistent with tracing outer structures because due to momentum
conservation rotating structures should have lower velocities further
out.

Notwithstanding that we do not exactly know the age of IRDC\,18223-3,
its non-detection up to 8\,$\mu$m puts it at an early evolutionary
phase prior to the better studied HMPOs and Hot Molecular Cores. Our
data clearly show that even at such early stages molecular outflows
and rotating structures perpendicular to that have been developed, and
it is likely that closer toward the core center, one will find a real
accretion disk. To investigate the latter in more detail, higher 
angular resolution observations of an optically thin dense gas tracer
are required.

\subsection{The HMPO disk candidate IRAS\,18089-1732}

As a more evolved massive disk candidate, we have studied intensely
over the last few years the HMPO IRAS\,18089-1732. This source at a
distance of 3.6\,kpc with a luminosity of $10^{4.5}$\,L$_{\odot}$ is
part of a large sample of HMPOs, it hosts H$_2$O and Class {\sc ii}
CH$_3$OH maser and has strong molecular line emission indicative of en
embedded Hot Molecular Core \citep{sridha,beuther2002a}. During early
SMA observations \citet{beuther2004a,beuther2004b,beuther2005c}
identified in SiO a molecular outflow in the north-south direction,
and perpendicular to that in HCOOCH$_3$ a velocity gradient on scales
of a few 1000\,AU. Although these data were indicative of rotation and
an underlying massive accretion disk, the observations did not allow
us to characterize the structure in more detail because of a lack of
spatial resolution. Therefore, we observed IRAS\,18089-1732 now in
high-energy transitions of NH$_3$ at 1.2\,cm wavelength with the VLA
and the ATCA at a spatial resolution of $0.4''$ \citep{beuther2008a}.
These NH$_3$(4,4) and (5,5) lines have a two-fold advantage: Their
high excitation levels ($>200$\,K) ensure that we are tracing the warm
inner regions and are less confused by the surrounding cold envelope,
whereas the cm wavelengths regime is less affected by high optical
depth of the dust emission in high column density regions and may
hence be particularly well suited for massive disk studies (e.g.
\citealt{krumholz2007a}). Figure \ref{18089} presents an integrated
image and a 1st moment map (intensity weighted velocity) of the
corresponding VLA observations.

\begin{figure}[htb] 
\begin{center}
\end{center}
\caption{\small The left panel shows the VLA NH$_3$(5,5) emission
  integrated from 31 to 37\,km\,s$^{-1}$ \citep{beuther2008a}. The
  right panel presents the corresponding 1st moment map contoured from
  31.5 to 36.5\,km\,s$^{-1}$ (step 1\,km\,s$^{-1}$).  The white-black
  dashed contours show the 1.2\,cm continuum emission. The asterisks
  mark the position of the submm continuum peak \citep{beuther2005c},
  and the synthesized beams are shown at the bottom-left (grey NH$_3$
  and dashed 1.2\,cm emission).} \label{18089}
\end{figure}

The 1st moment map confirms the previously assessed velocity gradient
in east-west direction perpendicular to the molecular outflow. The
NH$_3$ line-width FWHM toward the central core is 4.7\,km\,s$^{-1}$,
significantly broader than that of IRDC\,18223-3
(\S\ref{18223-cassie}). In the simple picture of equilibrium between
gravitational and centrifugal force, the rotationally bound mass would
be $\sim$37\,M$_{\odot}$, of the same order as the whole gas mass as well
as the mass of the central source (of the order 15\,M$_{\odot}$).
Furthermore, the position-velocity diagram is not consistent with
Keplerian rotation. It even shows indications of super-Keplerian
motion, which is expected for very massive disks where the rotation
profile is not only determined by the mass of the central object but
also by the disk itself (e.g., \citealt{krumholz2006b}). Hence, the
new VLA and ATCA data clearly confirm the previous assessment of
rotation perpendicular to the outflow/jet, however, the kinematic
signatures of that rotating structure are not consistent with a
Keplerian disk like in low-mass star formation, but they show
additional features which can be produced by massive self-gravitating
disks as well as by infalling gas that may settle eventually on the
disk. In addition to this, the detection of the high-excitation lines
in the rotating material indicates high average gas temperatures
$>$100\,K for the disk-like structures, well in excess of typical gas
temperatures in low-mass disks of the order 30\,K (e.g.,
\citealt{pietu2007}).  Moreover, we detect double-lobe cm continuum
emission close to the core center where the two lobes are oriented in
north-south direction parallel to the outflow identified in SiO. With
respect to previous data at longer wavelength, we find a spectral
index at cm wavelength of 1.9, consistent with an optically thick jet
\citep{reynolds1986}.

It will be interesting to further zoom into the innermost regions with
future instruments like ALMA and eVLA to asses whether the
quantitative deviations from typical low-mass accretion disks continue
down to the smallest scales, or whether we will find Keplerian disk
structures as known from their low-mass counterparts.

\subsection{A more evolved massive disk candidate?}

Moving along in the evolutionary sequence, we have recently identified a
potential disk around a more evolved candidate young stellar object
(Quanz et al. in prep.). The source labeled so far mdc1 (massive disk
candidate) was identified serendipitously during a near-infrared
wide-field imaging campaign on Calar Alto via its K-band cone-like
nebulosity and a central dark lane (Fig.~\ref{mdc1}). First
single-dish bolometer and spectral line measurements revealed a
1.2\,mm flux of 12\,mJy and a velocity of rest of
$\sim$51\,km\,s$^{-1}$. The latter value indicates a kinematic
distance of $\sim$5\,kpc, consistent with distances of a few UCH{\sc
  ii} regions in the surrounding neighborhood. To investigate this
object in more detail, we recently observed it with the SMA at
1.3\,mm wavelength mainly in the mm continuum and the
$^{12}$CO/C$^{18}$O spectral line emission. Figure \ref{mdc1}
presents an overlay of the SMA data with the K-band nebulosity, and a
few points need to be stressed:

\begin{figure}[htb] 
\begin{center}
\end{center}
\caption{\small The grey-scale in both panels shows the K-band
  near-infrared nebulosity observed for this massive evolved disk
  candidate (Quanz et al.~in prep.). The contours are the
  corresponding SMA mm observations where the left panel shows the
  1.3\,mm continuum emission, and the right panel in black and white
  contours the blue-shifted $^{12}$CO(2--1) and the integrated
  C$^{18}$O(2--1) emission, respectively.} \label{mdc1}
\end{figure}

(a) Although spatially unresolved with a synthesized beam of $\sim 4.0''$
the 1.3\,mm continuum peak exactly coincides with the infrared dark
lane consistent with the large column densities of the proposed
disk-like structure. The flux measured with the SMA is 12\,mJy
like the previous single-dish measurements. This indicates that we
have no surrounding dust/gas envelope but rather an isolated central
structure. Assuming optically thin dust emission at 50\,K, the
approximate gas mass of the central structure is $\sim$5\,M$_{\odot}$.
(b) We detect blue-shifted CO(2--1) spatially well correlated with the
K-band nebulosity north of the dark lane. This confirms the initial
interpretation of that feature to be due to an outflow. (c) The
integrated C$^{18}$O(2--1) emission is elongated perpendicular to the
outflow observed in CO and K-band continuum emission. The line-width
FWHM of the C$^{18}$O emission is narrow with $\sim$0.8\,km\,s$^{-1}$,
however, the spatial extent of this structure is large, of the order
$2\times10^4$\,AU.

While the low gas mass and the missing more massive gas envelope could
be interpreted in the framework of a low-mass source, such large
disk-structures as indicated by the C$^{18}$O emission are not known
from typical low-mass disk sources (e.g., \citealt{simon2000}).
Therefore, these observations can also be interpreted as a remnant
disk/torus around an intermediate to high-mass (proto)star that has
already dispersed much of its envelope. Although C$^{18}$O is detected
only in two channels, these show a clear velocity shift, and the small
line-width may be due to the lower rotational motions on large scales
assuming momentum conservation in rotating, potentially Keplerian
structures.\\

Synthesizing the three example sources shown here, it is
interesting to note that the line-widths are small in the youngest and
the supposed to be oldest source, whereas they are large in the HMPO
which should be in its main accretion phase. In an evolutionary
picture this can be interpreted that at early evolutionary stages
infall, turbulence and rotation are not yet that vigorous. Then in the
main accretion phase, infall, rotation and outflow processes strongly
increase the line-width. And finally, when the accretion stops, the
envelope and disk slowly disperse and one observes only a remnant
structure with small line-width in the outer regions. This scenario is
speculative, however, the number of disk candidates is steadily
increasing, and since we start sampling more evolutionary stages, we
are getting the chance to address disk evolution questions in
high-mass star formation as well.

\section{Fragmentation in high-mass star formation}

How massive gas clumps fragment is one of the key questions if one
wants to understand the formation of the Initial Mass Function, and as
outlined in \S \ref{intro}, the two main massive star formation theories
predict differences in the early fragmentation processes. In the
following I will present several examples of fragmenting massive cores
addressing issues about fragmentation on the cluster-scale,
fragmentation of smaller groups, potential proto-trapezia, and the
determination of density structures of sub-sources within evolving
clusters.

\subsection{Resolving the massive proto-cluster IRAS\,19410+2336}

To address fragmentation processes at early evolutionary stages high
angular resolution at (sub)mm wavelengths is the tool of choice to
resolve the relevant substructures. \citet{beuther2004c} resolved the
young massive star-forming region IRAS\,19410+2336 (distance
$\sim$2.1\,kpc and luminosity $\sim$10$^{4}$\,L$_{\odot}$) at 1.3\,mm
wavelength with the PdBI at approximately 2000\,AU linear resolution
into 24 sub-sources. Although from a statistical point of view such
numbers cannot compete with the clusters exceeding 100 or even 1000
stars observed at optical and near-infrared wavelength, this is still
one of the prime examples of a spatially resolved massive
proto-cluster. Assuming that all emission features are due to cold
dust emission from embedded protostars, they were able to derive a
core-mass function. With a power-law slope of -2.5, this core mass
function is consistent with the Salpeter IMF slope of -2.35
\citep{salpeter1955}. Therefore, \citet{beuther2004c} interpreted
these observations as support for the turbulent fragmentation put
forth by, e.g., \citet{padoan2002}.

A few caveats need to be kept in mind: While \citet{beuther2004c}
assumed a uniform gas temperature for all sub-sources, it is more
likely that the central peaks are warmer than those further outside.
This issue can be addressed by spectral line emission with
temperature sensitive molecules (e.g., H$_2$CO) which is an ongoing
project by Rodon et al.~(in prep.).  Furthermore, the assumption that
all mm continuum peaks are of pro- or pre-stellar nature is not
necessarily always valid, e.g., \citet{gueth2003} or
\citet{beuther2004d} have shown that mm continuum emission can partly
also be caused by collimated jets.  However, only the central source
is detected at cm wavelength and collimated jets should be detectable
at cm wavelengths as well. Therefore, we believe that jets should not
affect the analysis much. 

Independent of the caveats, it is surprising that IRAS\,19410+2336 is
still the only young massive star-forming region that is resolved in
$>$10 sub-sources in the mm continuum emission. While this can be
explained to some degree by the exceptionally good uv-coverage
obtained for the given observations, which results in a good sampling
of spatial structures, we also need to consider whether different
modes of fragmentation may exist. Similar high-spatial-resolution
studies of more proto-clusters spanning a broad range of luminosities
are required to tackle this question in more detail. Another
interesting question is associated with the spatial filtering of
interferometers and the corresponding large-scale emission: Many
interferometric (sub)mm continuum studies of massive star-forming
regions filter out of the order 90\% of the flux, hence, large amounts
of the gas are distributed on larger scales, usually $>10''$. The
question remains whether this gas will eventually participate in the
star formation process or not?

\subsection{Fragmentation of potential proto-trapezia}

\subsubsection{The enigmatic proto-trapezium W3-IRS5}

The W3-IRS5 region is one of the prototypical high-mass star-forming
regions with $\sim 10^5$\,L$_{\odot}$ at a distance of $\sim$1.8\,kpc
that shows fragmentation on scales of the order 1000\,AU observed at
near-infrared as well as cm wavelengths
\citep{megeath2005,vandertak2005a}. However, not much was known about
the cold dust and gas emission. Therefore, we observed the region with
the PdBI at 1.3 and 3.5\,mm wavelengths with the new extended
baselines resulting in an unprecedented spatial resolution of $\sim
0.37''$ (Rodon et al.~in prep.). Figure \ref{w3irs5} shows a
compilation of the 1.3\,mm continuum data and the SiO(5--4) and
SO$_2(22_{2,20}--22_{1,21}$) spectral line emission.

\begin{figure}[htb] 
\caption{\small PdBI observations of the W3-IRS5 system from Rodon et
  al. (in prep.). The left panel shows the 1.3\,mm continuum emission,
  and the stars mark near-infrared sources from \citet{megeath2005}.
  The middle panel presents as solid and dotted contours the blue- and
  red-shifted SiO(5--4) emission overlaid on the grey-scale 1.3\,mm
  continuum emission. The right panel finally shows the 1st moment map
  of SO$_2(22_{2,20}--22_{1,21}$ in grey-scale with the 1.3\,mm
  continuum contours.}
\label{w3irs5}
\end{figure}

The mm continuum emission resolves the W3-IRS5 region into five
sub-sources, where four of them are coincident with near-infrared and
cm emission peaks. Three of the sources are clustered in a very small
projected volume of only $\sim$2000\,AU. With this high spatial
resolution we find extremely large average column densities of the
order a few times $10^{24}$\,cm$^{-2}$ which corresponds to visual
extinctions $A_v$ between $5\times10^3$ and $10^4$ averaged over the
beam size. Such extinctions should be far too large to allow any
detection at near-infrared wavelengths, nevertheless, near-infrared
counterparts are detected \citep{megeath2005}. This conundrum can
likely be explained by the detection of several SiO outflows in the
field. In particular, we find very compact blue- and red-shifted SiO
emission toward the two main mm peaks, where the blue- and red-shifted
emission is barely spatially separated (Fig.~\ref{w3irs5} middle
panel). Since the overall time-scale of the W3-IRS5 outflow system is
relatively large (of the order a few times $10^4$\,yrs,
\citealt{ridge2001}), these compact features are unlikely from very
young outflows, but they indicate that the outflows are
oriented almost along the line of sight. The opening cones of the
outflows are the likely cause that emission from close to the
protostars can escape the region and hence make them detectable at
near-infrared wavelengths. 

The right panel of Fig.~\ref{w3irs5} shows the 1st moment map of
SO$_2$ (intensity weighted velocities) which encompasses the mm
continuum peaks. The coherent velocity field over the sub-sources is a
strong indicator that the system is a bound structure and not some
unbound chance alignment within the field (e.g.,
\citealt{launhardt2004}). In addition to the general velocity gradient
from the south-east to the north-west, one tentatively identifies
velocity gradients across the two strongest mm continuum peaks. Since
we do not know the exact orientation of the outflows with respect to
the SO$_2$ rotation axis, it is not yet possible to identify these
structures with disk-like components as in \S\ref{disks}. Future
observations in different tracers may help to shed more light on the
rotational structure associated with each sub-source. It should also
be noted that the line-width FWHM toward the mm continuum peaks varies
between 6.2 and 7\,km\,s$^{-1}$ larger than those reported in
\S\ref{disks}. While the larger line-width compared with the IRDC and
the more evolved source may be explained by the evolutionary sequence
sketched at the end of \S\ref{disks}, the larger FWHM compared to the
HMPO may have different reasons, among them are the larger luminosity
of W3-IRS5, its multiplicity compared with the so far unresolved
source IRAS\,18089-1732, and also the molecular species, because
SO$_2$ should be more affected by shocks than the NH$_3$ line used for
the IRAS\,18089-1732 study. In addition to this, the SO$_2$ moment map
exhibits a velocity discontinuity with a velocity jump of the order
4\,km\,s$^{-1}$ south-east of the mm continuum peaks. What is the
cause of this discontinuity, is it associated with the original core
formation and a shock within converging flows, or is it of
different origin?

In summary, the combination of high-spatial-resolution observations of
the continuum emission in addition to outflow and dense gas tracers
allows us to characterize many physical properties of this
proto-trapezium system with respect to its multiple components and
their outflow and rotation properties.

\subsubsection{Fragmentation of the hot core G29}

The hot core G29.96 located right next to a well-known cometary H{\sc
  ii} region comprises another example of several protostellar submm
continuum sources within the innermost center of a high-mass star
forming region (distance $\sim$6\,kpc, luminosity $9\times
10^4$\,L$_{\odot}$). High-spatial resolution observations with the SMA
in its most extended configuration yielded a spatial resolution of
$0.36''\times 0.25''$ in the submm continuum at $\sim$348\,GHz,
corresponding to a linear resolution of 2000\,AU (Fig.~\ref{g29},
\citealt{beuther2007d}, the line data will be discussed in
\S\ref{sequence}). The Hot Molecular Core previously identified in a
high-excitation NH$_3$ line \citep{cesaroni1998} is resolved by these
new data into four sub-sources within a projected diameter of
$\sim$6900\,AU. Assuming that the emission peaks are of protostellar
nature, \citet{beuther2007d} estimated a protostellar density of $\sim
2\times 10^5$\,protostars/pc$^{-3}$. This is considered a lower limit
since we are limited by spatial resolution, sensitivity and projection
effects. Nevertheless, such a protostellar density is about an order
of magnitude higher than values usually reported for star-forming
regions (e.g., \citealt{lada2003}). Although this value is still about
an order of magnitude lower than protostellar densities that would be
required in the merging scenario for massive stars (e.g.,
\citealt{bonnell2004,bally2005}), it is interesting to note that
increasingly higher protostellar densities are reported when going to
younger sources and better angular resolution (see also
\citealt{megeath2005}). This allows us to speculate whether future
observations with better spatial resolution and sensitivity toward
extremely massive star-forming regions will reveal protostellar
densities that may be sufficient to make mergers possible.  While such
a detection would not be a proof for mergers to exist, it will
certainly be important to verify whether the required initial
conditions do exist at all.

\begin{figure}[htb] 
\begin{center}
\end{center}
\caption{\small Compilation of data toward the UCH{\sc ii}/hot core
  region G29.96 from \citealt{beuther2007d}. The dashed contours
  present the cometary UCH{\sc ii} regions whereas the full contours
  show the older NH$_3$ observation from the hot core
  \citep{cesaroni1994}. The grey-scale with contours then present the
  new high-resolution ($0.36''\times 0.25''$) submm continuum data
  from the SMA.}
\label{g29}
\end{figure}

\subsection{Density structure of sub-sources -- IRAS\,05358+3543}

As a final example for the potential of (sub)mm continuum studies, I
present the recent multi-wavelength investigation of the HMPO
IRAS\,05358+3543. This region at a distance of 1.8\,kpc with a
luminosity of $10^{3.8}$\,L$_{\odot}$ was observed in a combined
effort with the PdBI and the SMA at arcsecond resolution in four
wavelength bands (3.1 and 1.2\,mm, and 875 and 438\,$\mu$m,
\citealt{beuther2007c,leurini2007}). While many details about the sub-structure of
the forming cluster can be derived, here, I will discuss only two
results.

Based on the multi-wavelength data, \citet{beuther2007c} fitted the
spectral energy distribution on the Rayleigh-Jeans side of the
spectrum (Fig. \ref{sed}). While the main source can well be fitted by
a typical protostellar spectrum consisting of free-free emission at
long wavelength and a steep flux increase at shorter wavelength due to
the dust emission, another sub-source did not fit at all into that
picture. In particular the shortest wavelength data-point at
438\,$\mu$m shows significantly lower fluxes than expected for a
typical protostar. The most likely explanation for this effect is that
we are dealing with a very cold source and that therefore we are
already approaching the peak of the spectral energy distribution. The
data allowed us to estimate an upper limit for the dust temperature of
$\leq 20$\,K. Since we are also not detecting any other line emission
from this core (mainly from typical hot core molecules,
\citealt{leurini2007}), it may well be a starless core right in the
vicinity of an already more evolved massive protostar.  Further
investigations of this sub-source in typical cold gas tracers like
N$_2$H$^+$ or NH$_3$ are required to test this proposal.  Independent
of whether this source harbors an embedded protostar or not, these
observations show the importance of short wavelength data at high
spatial resolution if one wants to differentiate between critical core
parameters like the dust temperature.

\begin{figure}[htb] 
\caption{\small The left panel presents the SED toward the coldest
  sub-source in IRAS\,05358+3543 \citep{beuther2007c}. The parameters
  of the fits are marked in the figure. The right panel shows
  intensities averaged in uv-annuli and plotted versus the
  baseline-length for different sub-sources and wavelengths. Most can
  be well fitted by power-law distributions.}
\label{sed}
\end{figure}

Another physical parameter which has so far not been observationally
constrained for massive star formation, is the density profile of
individual sub-sources. While density profiles of low-mass
star-forming cores have well been characterized (e.g.,
\citealt{motte1998,wardthompson1999,andre2000}), in high-mass star
formation, density profiles were until now only derived with
single-dish observation covering scales of the whole cluster but not
individual sub-sources
(e.g., \citealt{beuther2002a,mueller2002,hatchell2003}). This is partly
due to the technical problem of interferometer observations that
filter out large amounts of the flux and hence make density profile
determinations from their images extremely unreliable. To overcome
this problem, \citet{beuther2007c} analyzed the data directly in the
uv-domain prior to any fourier transformation. Figure \ref{sed} shows
the corresponding plots of the observed intensities versus the
uv-distance for three sub-sources in three wavelengths bands,
respectively.  The observations cannot be fitted with Gaussian
distributions, but much better fits are achieved with power-law
distributions. These power-laws in the uv-domain can directly be
converted to the corresponding power-laws of the intensity profiles in
the image plane. Assuming furthermore a temperature distribution
$T\propto r^{-0.4}$ we can now infer the density profiles of
individual sub-sources of the evolving cluster. The derived
density profiles $\rho\propto r^{-p}$ have power-law indices $p$
between 1.5 and 2. Although this result is similar to the density
profiles previously determined for low-mass cores, to our knowledge
this is the first time that they have been observationally constrained
for resolved sub-sources in a massive star-forming region. The density
structure is an important input parameter for any model of star
formation (e.g., \citealt{mckee2003}).

\section{Astrochemistry}

 \subsection{Toward a chemical evolutionary sequence}
\label{sequence}

\begin{figure}[htb]
\includegraphics[angle=-90,width=5.9cm]{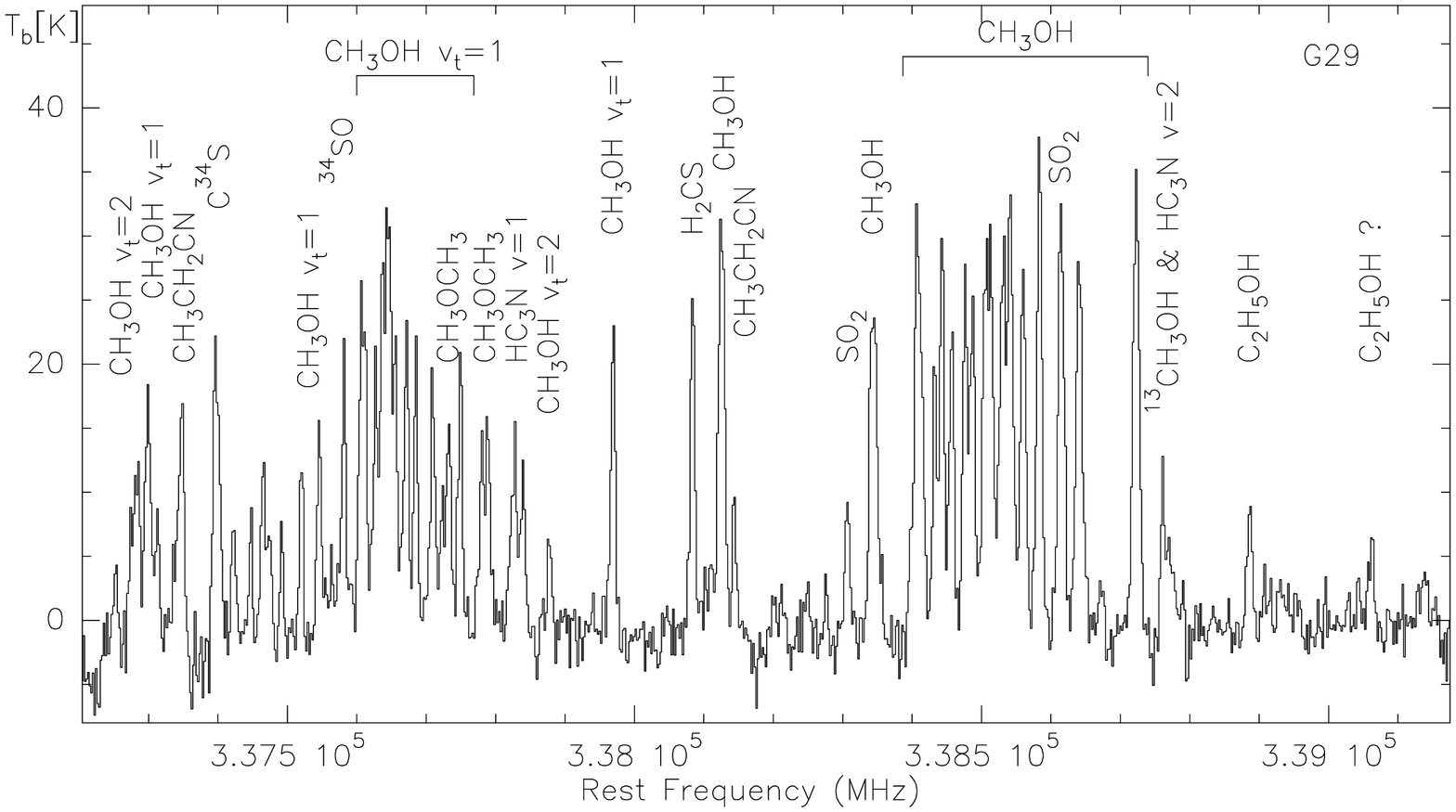}
\includegraphics[angle=-90,width=5.9cm]{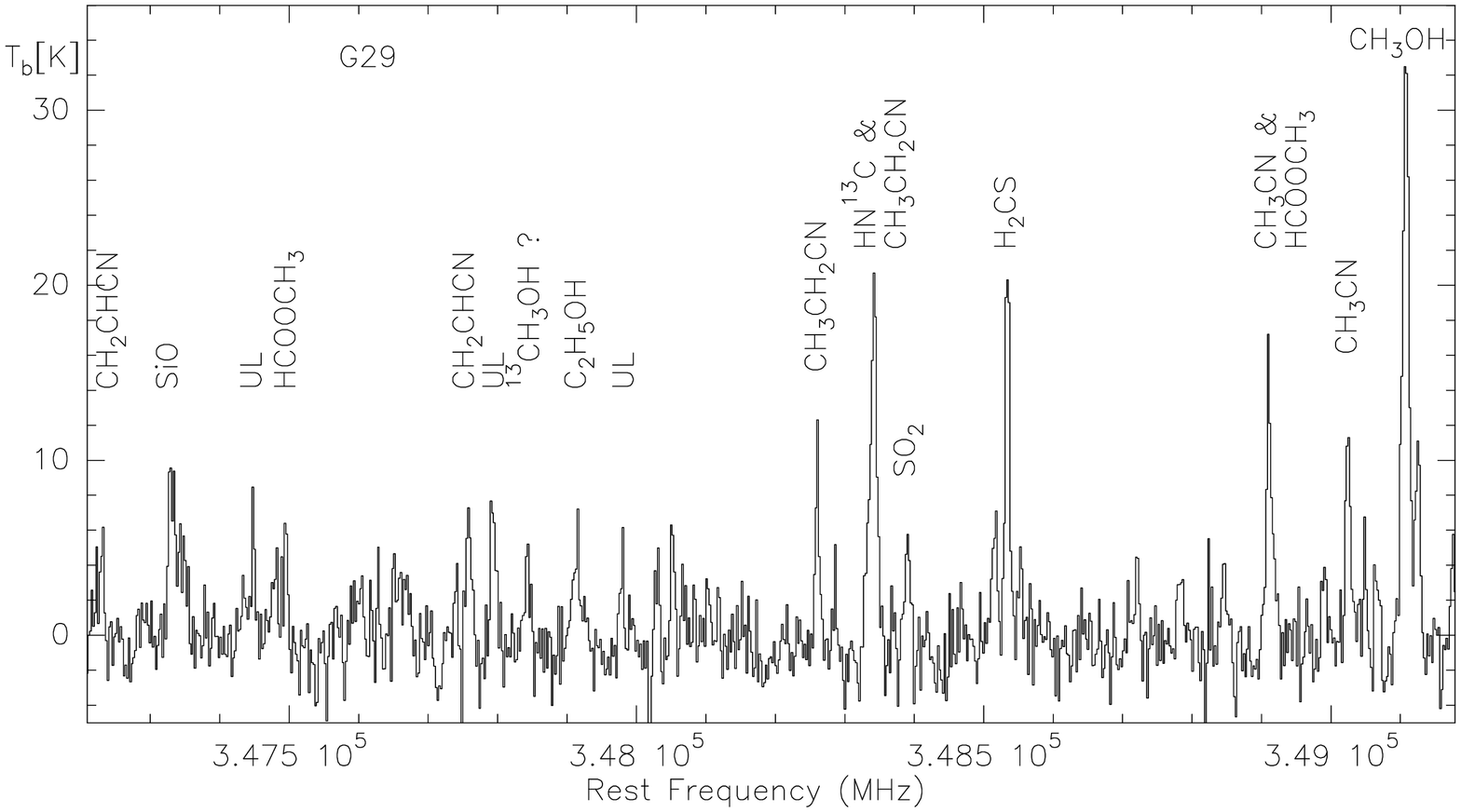}\\
\includegraphics[angle=-90,width=5.9cm]{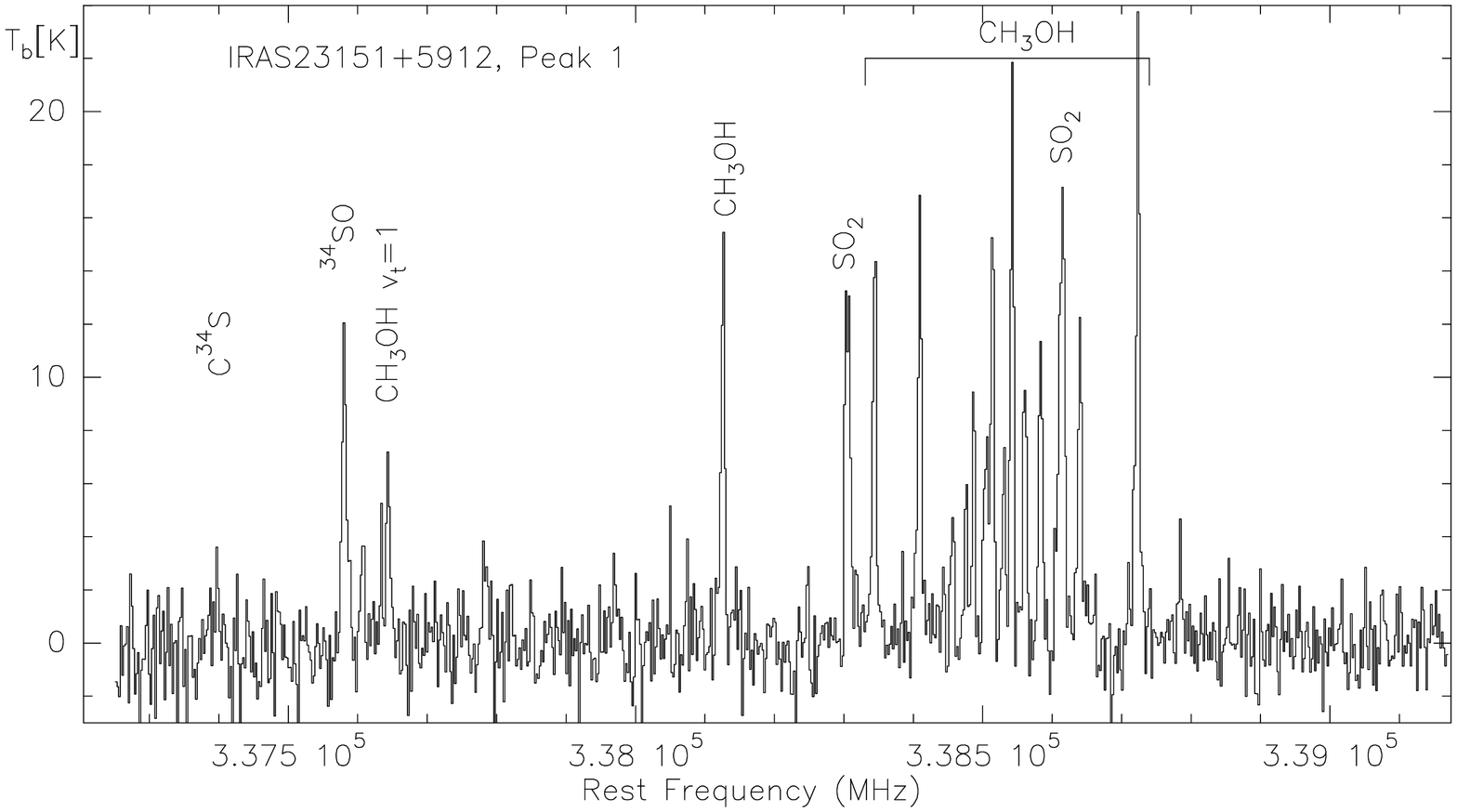}
\includegraphics[angle=-90,width=5.9cm]{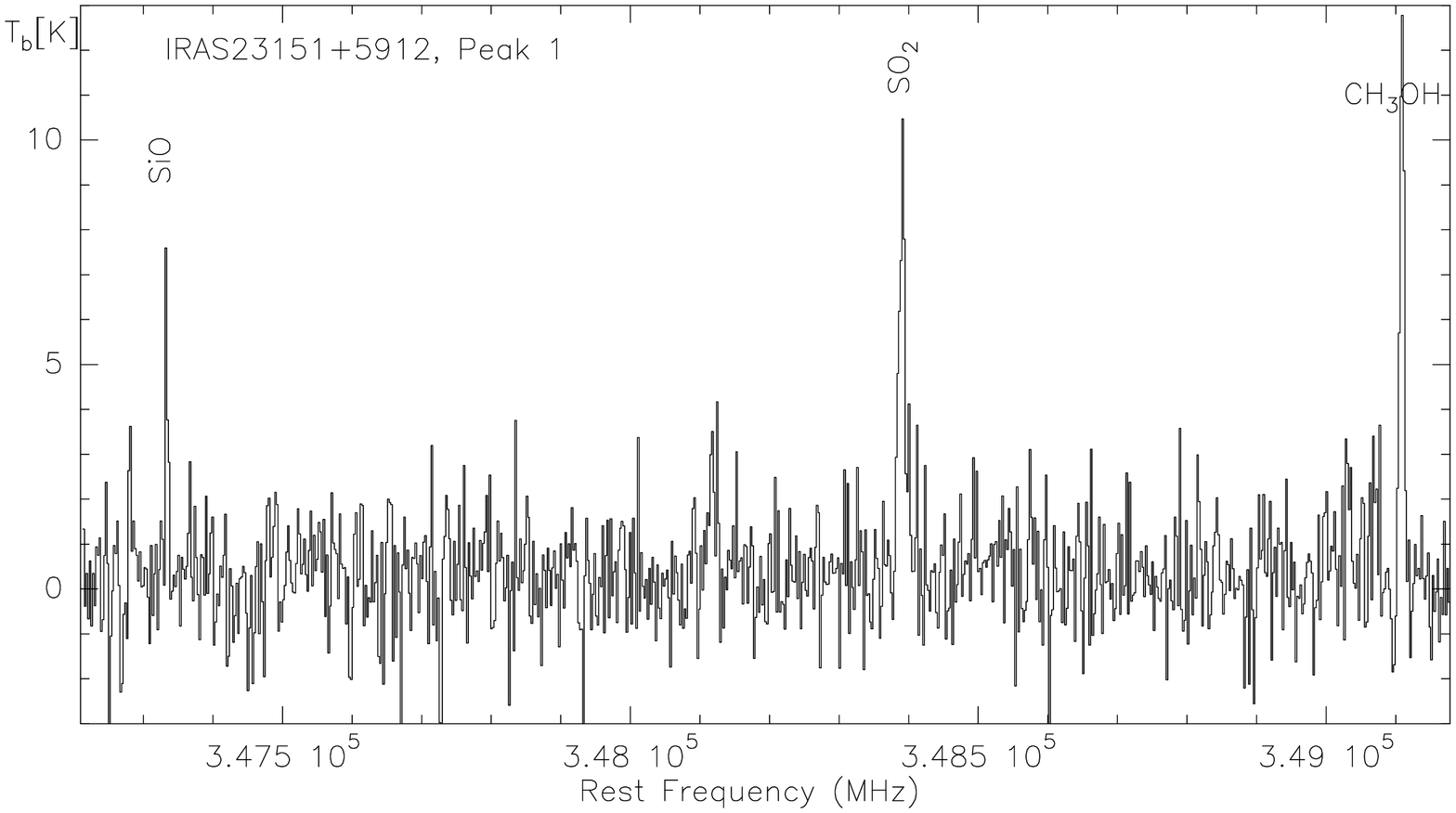}\\
\caption{\small SMA spectra extracted toward two massive star-forming
  regions (G29.96 top row \& IRAS\,23151+5912 bottom row, Beuther et
  al., subm.). The spectral resolution in all spectra is 2\,km/s. The
  left and right column show the lower and upper sideband data,
  respectively.}
\label{sample_spectra}
\end{figure}

Astrochemistry is a continuously growing field in astronomy. Although
line-survey style studies of different sources have existed for quite
some time (e.g., \citealt{blake1987,schilke1997b}), these studies had
usually been performed with single-dish instruments averaging the
chemical properties over the whole cluster-forming regions. Since the
advent of broadband receivers at interferometers like the SMA, it is
now also possible to perform imaging spectral line surveys that allow
us to spatially differentiate which molecules are present in which
part of the targeted regions, for example, the spatial differentiation
between nitrogen- and oxygen-bearing molecules in Orion-KL (e.g.,
\citealt{blake1996,beuther2005a}). In addition to the spatial analysis
of individual regions, we are also interested in analyzing how the
chemistry evolves in an evolutionary sense. As an early step in this
direction we synthesized SMA observations that were observed in the
same spectral setup around 862\,$\mu$m toward four massive
star-forming regions over the last few years (Beuther et al. subm.).
These four regions comprise a range of luminosities between
$10^{3.8}$\,L$_{\odot}$ and $10^5$\,L$_{\odot}$, and they cover
different evolutionary stages from young High-Mass Protostellar
Objects (HMPOs) to typical Hot Molecular Cores (HMCs): Orion-KL: HMC,
$L\sim 10^5$\,L$_{\odot}$, $D\sim 0.45$\,kpc \citep{beuther2005a};
G29.96: HMC, $L\sim 9\times 10^4$\,L$_{\odot}$, $D\sim 6$\,kpc
\citep{beuther2007d}; IRAS\,23151, HMPO, $L\sim 10^5$\,L$_{\odot}$,
$D\sim 5.7$\,kpc \citep{beuther2007f}; IRAS\,05358: HMPO, $L\sim
10^{3.8}$\,L$_{\odot}$, $D\sim 1.8$\,kpc
\citep{beuther2007c,leurini2007}. Smoothing all datasets to the same
linear spatial resolution of 5700\,AU, we are now capable to start
comparing these different regions. Figure \ref{sample_spectra}
presents typical spectra extracted toward the HMC G29.96 and the HMPO
IRAS\,23151.

A detailed comparison between the four sources is given in a
forthcoming paper (Beuther et al.~subm.), here we just outline a
few differences in a qualitative manner.\\
$\bullet$ The HMCs show far more molecular lines than the HMPOs.
Orion-KL and G29.96 appear similar indicating that the nature of the
two sources is likely to be comparable as well.  Regarding the two
HMPOs, the higher luminosity one (IRAS\,23151) shows still more lines
than the lower-luminosity source (IRAS\,05358). Since IRAS\,05358 is
approximately three times closer to us than IRAS\,23151, this is not a
sensitivity issue but it is likely due to
the different luminosity objects forming at the core centers.\\
$\bullet$ The ground-state CH$_3$OH lines are detected toward all four
sources.  However, the vibrational-torsional excited CH$_3$OH are only
strongly detected toward the HMCs Orion-KL and G29.96. Independent of
the luminosity, the HMPOs exhibit only one CH$_3$OH $v_t=1$ line,
which can easily be explained by the lower average temperatures of the
HMPOs.\\
$\bullet$ A more subtle difference can be discerned by comparing the
SO$_2$ and the HN$^{13}$C/CH$_3$CH$_2$CN line blend near 348.35\,GHz
(in the upper sideband). While the SO$_2$ line is found toward all
four sources, the HN$^{13}$C/CH$_3$CH$_2$CN line blend is strongly
detected toward the HMCs, but it is not found toward the HMPOs. In the
framework of warming up HMCs, this indicates that nitrogen-bearing
molecules are either released from the grains only at higher
temperatures, or they are daughter molecules which need some time
during the warm-up phase to be produced in gas-phase chemical
networks. In both cases, such molecules are not expected to be found
much prior to the formation of a detectable
HMC.\\
$\bullet$ Comparing the spatial distribution of different molecules,
we find, e.g., that C$^{34}$S is observed mainly at the core edges and
not toward the submm continuum peak positions. This difference can be
explained by temperature-selective desorption and successive gas-phase
chemistry reactions: CS desorbs early from the grains at temperatures
of a few 10\,K and should peak during the earliest evolutionary phases
toward the main continuum sources. Subsequently when the core warms up
to $\sim$100\,K, H$_2$O desorbs and dissociates to OH. The OH then
quickly reacts with the sulphur to form SO and SO$_2$ which should
then peak toward the main continuum sources. This is what we observe
in our data. The fact that the C$^{34}$S peaks are offset from the
submm continuum condensations even toward the younger sources is due
to their evolutionary stage where they have already heated up their
central regions to more than 100\,K.  Even younger sources are
required to confirm this scenario.

\subsection{C$_2$H as a tracer of the earliest evolutionary stages?}
\label{sec_c2h}

In an effort to study a larger source sample with respect to its
chemical evolution, we observed 21 massive star-forming regions
covering all evolutionary stages from IRDCs via HMPOs/hot cores to
UCH{\sc ii} with the APEX telescope at submm wavelengths (Beuther et
al.~subm.). While most spectral lines were detected mainly toward the
HMPO/hot core sources, the ethynyl molecule C$_2$H is omni-present
toward all regions. To get an idea about the spatial structure of
ethynyl, we went back to an older SMA data-set targeting the HMPO
IRAS\,18089-1732 at the same frequency around 349.4\,GHz of the C$_2$H
line \citep{beuther2005c}. Because we were not able to image the
spatial distribution of C$_2$H at that time, we now restricted the
data to only the compact configuration allowing us to better image the
larger-scale distribution of the gas. Figure \ref{c2h} presents the
resulting molecular line map, and we find that C$_2$H is distributed
in a shell-like fashion around the central protostellar condensation.
Comparing this with all other imaged molecules in the original paper,
only C$_2$H exhibits this behavior. To better understand this effect,
we ran a set of chemical models in 1D for a cloud of
1200\,M$_{\odot}$, a density power-law $\rho\propto r^{-1.5}$ and
different temperature distributions $T\propto r^q$. A snapshot of
these models after an evolutionary time of $5\times 10^4$\,yrs is
presented in Figure \ref{c2h}.  The models reproduce well the
central C$_2$H gap in IRAS\,18089-1732 which should have approximately
the same age.

\begin{figure}[htb]
\caption{\small The left panel presents in grey-scale the C$_2$H
  emission and in thick solid contours the corresponding submm
  continuum from the SMA toward the HMPO IRAS\,18089-1732 (Beuther et
  al., subm.). The right panel shows a chemical model explaining the
  decreased emission toward the core center after approximately
  $5\times 10^4$\,yrs. The parameter $q$ denotes the temperature
  power-law index, and the $T$ values refer to the temperature at the
  core edge or to isothermal values ($q=0$).}
\label{c2h}
\end{figure}

While these models reproduce the observations, they give also
predictions how the C$_2$H emission should look like at different
evolutionary times. In particular, C$_2$H forms quickly early on, also
at the core center. Since not many molecules exist which do not freeze
out and are available to investigate the cold early phases of massive
star formation (valuable exceptions are, e.g., NH$_3$ or N$_2$H$^+$),
the detection of C$_2$H toward the whole sample in combination with
the chemical models triggers the prediction that C$_2$H may well be an
excellent molecule to investigate the physical conditions of (massive)
star-forming regions at very early evolutionary stages.
High-spatial-resolution observations of IRDCs are necessary to
investigate this potentially powerful astrophysical tool in more
detail.

\subsection{Employing molecules as astrophysical tools}

While the chemical evolution of massive star-forming regions is
interesting in itself, one also wants to use the different
characteristics of molecular lines to trace various physical
processes. In contrast to molecules like SiO and CO that are
well-known outflow/jet tracers, the task gets more difficult searching
for suitable accretion disk tracers. Investigating our sample and disk
claims in the literature, one finds that in many cases exclusively one
or the other molecule allows the investigation of rotational motion,
whereas most other molecular lines remain without clear signatures.
For example, the HN$^{13}$C line discussed above (\S \ref{sequence})
traces rotation in the hot core G29.96 but it is not even detectable
in younger sources. The other way around, C$^{34}$S traced disk
rotation in the young HMPO IRAS\,20126 \citep{cesaroni2005}, but not
anymore toward more evolved sources (\S\ref{sequence}).  These
differences imply that one will unlikely find a uniquely well suited
molecular line allowing the study of large samples of massive
accretion disks, but that one has to select for each source or source
class the suitable molecule for detailed investigations.

In the following, I give a short table with molecules and their
potential usefulness to study different physical processes. This table
(1) is restricted to molecules with spectral lines at cm/(sub)mm
wavelengths and does not claim any kind of completeness, it should
just serve as a rough overview and it only lists the main
isotopologues of each species.

\begin{table}[htb]
\begin{tabular}{ll}
\hline
OH & Zeeman effect, magnetic fields, maser signpost of star\\
       &  formation  \\
CO & General cloud structure, outflows \\
SiO & Shocks due to jets/outflows\\
CO$^+$ & Far-UV radiation from embedded protostars \\
CS & Dense gas, rotation, also outflows \\
CN & Photodominated regions, Zeeman effect, magnetic fields \\
SO & Shocks, dense gas \\
H$_2$O & Shocks and hot cores, rotation (H$_2^{18}$O), maser signpost of\\
                 & star formation \\
HDO & Deuterium chemistry \\
H$_2$D$^+$ & Cold gas, pre-stellar cores, freeze out \\
HCN & Dense cores, also outflows \\
HNC & Dense cores, rotation (HN$^{13}$C) \\
HCO$^+$ & Outflows, infall, cosmic rays, ionization degree, \\
                   & dense gas (H$^{13}$CO$^+$) \\
SO$_2$ & Shocks, dense gas \\
C$_2$H & Early evolutionary stages (\S\ref{sec_c2h})\\
N$_2$H$^+$ & Early evolutionary stages \\
N$_2$D$^+$ & Deuteration, freeze out \\
H$_3$O$^+$ & Cosmic rays \\
H$_2$CO & Dense gas, temperatures \\
NH$_3$ & Cold and hot cores, rotation, temperatures \\
CH$_3$OH & Shocks, young rotating structures? (\S\ref{18223-cassie}), temperatures, \\
                       & maser signpost of massive star formation \\
CH$_3$CN & Hot cores, temperatures, rotation \\
CH$_3$CCH & Dense gas, temperatures \\
HCOOCH$_3$ & Hot cores, rotation \\
CH$_3$CH$_3$CN & Hot Cores \\
\hline
\end{tabular}
\label{linelist}
\caption{A few useful molecules and some of their potential applications.}
\end{table}

\section{Conclusions and summary}

This article tries to outline how far we can currently constrain
physical and chemical properties in massive star formation using
(sub)mm interferometry. Coming back to the original questions raised
in the abstract: (a) What are the physical conditions at the onset of
massive star formation? (b) What are the characteristics of potential
massive accretion disks and what do they tell us about massive star
formation in general? (c) How do massive clumps fragment, and what
does it imply to high-mass star formation? (d) What do we learn from
imaging spectral line surveys with respect to the chemistry itself as
well as for utilizing molecules as tools for astrophysical
investigations?
  
Can we reasonably answer any of these questions with confidence? There
are no clear-cut answers possible yet, however, the observations are
paving a way to shedding light on many of the issues, and one can try
to give tentative early answers. The following is a rough attempt to
outline the directions for current and future answers in these fields:

(a) Massive gas clumps prior or at the onset of high-mass star
formation are characterized by cold temperatures of the order 15\,K
and small line-widths indicative of a low level of turbulence. Their
molecular abundances appear comparable to those of low-mass starless
cores. Interestingly, the outflow detection rates toward IRDCs are
high, and no genuine High-Mass Starless Cores have been reported in
the literature yet. Although the statistical basis is not solid enough
yet, this allows us to speculate that the high-mass starless phase is
likely to be very shortlived.

(b) The detection of a real accretion disk around a massive protostar
still remains an open issue. However, we find many rotating structures
in the vicinity of young massive star-forming regions all the way from
IRDCs to Hot Molecular Cores. These structures are on average large
with sizes between $1\times 10^3$ and $2\times 10^4$\,AU, and they
have masses of the order of the central protostar. Hence, most of them
are not Keplerian accretion disks but rather some larger-scale
rotating/infalling structures or toroids that may feed more genuine
accretion disks in the so far unresolved centers of these regions.

(c) Fragmentation of massive star-forming regions is frequently
observed, and the core mass function of one young region is consistent
with the Initial Mass Function. However, caveats of unknown
temperature distributions or missing flux on larger scales may still
affect the results. Furthermore, we find proto-trapezium like structures
which show multiple bound sources on small scales of a few 1000\,AU
implying protostellar densities of the order $10^5$
protostars/pc$^{-3}$. Such densities are still not sufficient to allow
coalescence, however, it may be possible to find even higher
protostellar densities with the improved observational capabilities of
future instruments. Although mergers do not appear necessary to form
massive stars in general, they still remain a possibility for the most
massive objects.

(d) Astro-chemistry is a young branch in astrophysical research,
and we are currently only touching the surface of its potential. The
different paths to follow in the coming years are manyfold: With
larger source-samples, we will be able to derive a real chemical
evolutionary sequence with one of the goals to use chemistry as an
astrophysical clock. Furthermore, understanding the chemical
differences is important to use the molecular lines as astrophysical
tools to investigate the physical processes taking place. Moreover,
another current hot topic is planet formation, and in this context
astro-biology is a rising subject. In this regard understanding
astro-chemistry and detecting new and more complex molecules in space
is paving the way for future astro-biological science.

\noindent{\small {\bf Acknowledgments:} Thanks a lot to Cassie
  Fallscheer and Javier Rodon for preparing the figures related to the
  IRDC\,18223-3 outflow/disk system and the W3-IRS5 fragmenting core.
  I further acknowledge financial support by the Emmy-Noether-Program
  of the Deutsche Forschungsgemeinschaft (DFG, grant BE2578).}

\input{refs}

\input{biermann_beuther.bbl}





\vfill

\end{document}

%% file: refs.tex
\def\aj{AJ}%
\def\araa{ARA\&A}%
\def\apj{ApJ}%
\def\apjl{ApJ}%
\def\apjs{ApJS}%
\def\ao{Appl.~Opt.}%
\def\apss{Ap\&SS}%
\def\aap{A\&A}%
\def\aapr{A\&A~Rev.}%
\def\aaps{A\&AS}%
\def\azh{AZh}%
\def\baas{BAAS}%
\def\jrasc{JRASC}%
\def\memras{MmRAS}%
\def\mnras{MNRAS}%
\def\pra{Phys.~Rev.~A}%
\def\prb{Phys.~Rev.~B}%
\def\prc{Phys.~Rev.~C}%
\def\prd{Phys.~Rev.~D}%
\def\pre{Phys.~Rev.~E}%
\def\prl{Phys.~Rev.~Lett.}%
\def\pasp{PASP}%
\def\pasj{PASJ}%
\def\qjras{QJRAS}%
\def\skytel{S\&T}%
\def\solphys{Sol.~Phys.}%
\def\sovast{Soviet~Ast.}%
\def\ssr{Space~Sci.~Rev.}%
\def\zap{ZAp}%
\def\nat{Nature}%
\def\iaucirc{IAU~Circ.}%
\def\aplett{Astrophys.~Lett.}%
\def\apspr{Astrophys.~Space~Phys.~Res.}%
\def\bain{Bull.~Astron.~Inst.~Netherlands}%
\def\fcp{Fund.~Cosmic~Phys.}%
\def\gca{Geochim.~Cosmochim.~Acta}%
\def\grl{Geophys.~Res.~Lett.}%
\def\jcp{J.~Chem.~Phys.}%
\def\jgr{J.~Geophys.~Res.}%
\def\jqsrt{J.~Quant.~Spec.~Radiat.~Transf.}%
\def\memsai{Mem.~Soc.~Astron.~Italiana}%
\def\nphysa{Nucl.~Phys.~A}%
\def\physrep{Phys.~Rep.}%
\def\physscr{Phys.~Scr}%
\def\planss{Planet.~Space~Sci.}%
\def\procspie{Proc.~SPIE}%
\let\astap=\aap
\let\apjlett=\apjl
\let\apjsupp=\apjs
\let\applopt=\ao